\documentclass[graybox]{svmult}

\usepackage{type1cm}        
\usepackage{makeidx}         
\usepackage{graphicx}        
\usepackage{sidecap}          
           
\usepackage{multicol}        
\usepackage[bottom]{footmisc}

\usepackage{url}

\usepackage{newtxtext}       %
\usepackage{newtxmath}       

\usepackage{tikz}
\usepackage{mathtools}

\makeindex             
                       
\newcommand{\be}{\nopagebreak[3]\begin{equation}}
\newcommand{\ee}{\end{equation}}
\newcommand{\ba}{\nopagebreak[3]\begin{eqnarray}}
\newcommand{\ea}{\end{eqnarray}}

\newcommand{\bc}{}

\newcommand{\myarrow}[1][-45]{%
  \mathrel{%
    \text{$
     \begin{tikzpicture}[baseline = + 0.9ex]
       \node[inner sep=0pt,outer sep=0pt,rotate = #1] (a) at (0,0)  {$\longrightarrow{}$};
    \end{tikzpicture}
    $}%
  }%
}%

\begin{document}

\title*{The relational ontology of contemporary physics}
  \titlerunning{The relational ontology of contemporary physics} 
\author{Francesca Vidotto}
\institute{Francesca Vidotto \at The University of Western Ontario\\ \emph{Department\,of\,Physics\,and\,Astronomy, Department\,of\,Philosophy, Rotman\,Institute\,of\,Philosophy}\\ \email{fvidotto@uwo.ca}\\[1em]
{\it This is a preprint of the following chapter: Francesca Vidotto, The relational ontology of contemporary physics, to be published in ``Quantum Mechanics and Fundamentality: Naturalizing Quantum Theory between Scientific Realism and Ontological
 Indeterminacy'', edited by Valia Allori, 2022, Springer, reproduced with permission of Springer. 
}
}
%
%
\maketitle

\abstract{Quantum theory can be understood as pointing to an ontology of relations.  I observe that this 
reading of quantum mechanics is supported by the ubiquity of relationality in contemporary fundamental
physics, including in classical mechanics, gauge theories, general relativity, quantum field theory, and the tentative theories
of quantum gravity. }


\section{Introduction}

The image of the world offered by quantum mechanics leaves us with a dizziness.  It is hard to make sense of the discreteness and the indetermination revealed by the theory. Different reactions to this vertigo lead to different ways of understanding what the theory tells us about reality, namely different 'interpretations' of the theory.  

Two opposite attitudes are possible. One is to try to bend discreteness and indetermination into an underlying hypothetical continuous and determined  reality.  For instance, the Many-Worlds interpretation assigns ontological value to the quantum states, which are continuous and always determined; while the DeBroglie-Bohm, or pilot-wave, interpretation assigns  ontological value to a commuting algebra of preferred variables such as positions of particles, which are also assumed to be continuous and always determined. 
All these are interpretations based on an ontology of objects, or entities, with properties.  

The other possible attitude is to take discreteness and indetermination at their face value, and study their consequences. The relational interpretation of quantum mechanics \cite{Rovelli:1995fv} starts from this second position.  It does not make use of an ontology of entities that have always properties, but rather an ontology of relations, where the \emph{properties} (of relata) are only determined at discrete interaction times and are always relative to both interacting systems\footnote{
The relationality of relational quantum mechanics has been compared with ontic structuralism by Candiotto and Dorato \cite{Candiotto:2017,Dorato:2020}. The metaphysical implications of relational quantum mechanics, and the association with the more general structuralist framework, are still requiring an in-depth investigation. In this respect, of particular interest are the positions developed by Esfeld, French and Ladyman \cite{Esfeld:2004,FrenchLadyman:2003}.
}.

This is a rather radical metaphysical position: it places relations,  rather than objects, or substances, at the center of the metaphysical conception. Articulating such a position has its difficulties: How to think of relations before relata? How to preserve the objectivity of our representations, if properties turn out to be so relative?  What grants the commensurability between perspectives? There are  answers to these questions \cite{
FrenchKrause:2006,LadymanRoss:2007,Rovelli:2017sky}, but they involve a radical rethinking of the conceptual basis of all our representations of reality.  So, why should we venture into this arduous journey, when more pacifying readings of quantum theory, compatible with a more naive realism, are available?

After all, different interpretations offer coherent frameworks for understanding the content of quantum mechanics and interpreting reality around us. Internal coherence is a necessary condition for a consistent interpretation of quantum mechanics, but is insufficient in helping us choosing between different interpretations.

One possibility to settle this problem is to delegate the answer to the future: some interpretations may turn out to more fruitful.  This is for instance how the debate on wether or not it is better to consider the Earth to be the center of the Universe (a non empirical question!) was settled: one option turned out to be definitely more fruitful. It has been argued that quantum gravity might be easier within one interpretation.  Or perhaps a future theory superseding quantum mechanics will require one particular interpretation \cite{Smolin,Valentini:2021izg}.  In all these cases, however, the new results against which to evaluate current interpretations are not yet available. 

Here I want, instead, to investigate a different strategy for evaluating interpretations of quantum theory: their coherence with the conceptual frameworks of {\em other} physical theories that best capture our recent advances in understanding the physical world.  I argue below that the relationality that characterizes the relational interpretation of quantum mechanics is in fact not so unconventional after all. Rather, it characterizes modern physics.  My aim is to provide in this way a more solid foundations to the idea that relationality is central to quantum mechanics: through the analysis of how relationality is present, perhaps  in a transversal way, in virtually all aspects of contemporary physics.

In fact, I shall argue that the relationality at the base of quantum mechanics is already present in classical mechanics: by putting this classical relationality in evidence, we  better situate the emergence of the more subtle  relationality of the quantum case. To this end, we need to look at classical theory from a modern perspective, in particular using the language of symmetries and gauge theories. This allows us to create a natural bridge with quantum mechanics in its relativistic version. When we talk about interpretations of quantum mechanics, it is misleading to restrict us to the non relativistic domain: we must consider the compatibility with quantum field theory, with Yang Mills theories and with gravity.   Relationality offers a key to do so. 

I also discuss the specific connection between quantum theory and the relativistic theory of the gravitational field. The problematic nature of this connection can be solved by using a common language: that of totally constrained systems. This can serve as common ground for understanding the foundational problems of quantum mechanics, gravity, and the role of symmetries/gauge, within a common conceptual framework.

If these steps are carried out carefully, then the image of a fundamental conceptual structure for understanding reality at our present level of knowledge opens up: that of covariant quantum fields. This is what quantum gravity is about:  a quantum description of the gravitational field must follow from a covariant description of quantum fields in full generality.  Physics forces us towards engaging metaphysically in a specific direction: everything that exists is quantum, everything that exists is covariant.  I argue below that this is clarified by seeing that everything quantum is relational, and everything covariant is relational. \\[1em]

\begin{center}
  \begin{tabular}{@{} rcl @{}}
    \hline \\
    ~particles + fields &   & ~~spacetime \\[.7em]
{ \bf Quantum Theory} ~ $\Downarrow$  \hspace{8mm} &  &  \hspace{6mm} $\Downarrow$ ~ { \bf General Relativity} \\[.5em]
    quantum fields & ~$\myarrow[-45]$ \hskip2cm $\myarrow[-135]$~ & covariant field \\
   & covariant quantum fields &\\[2em]
    \hline
  \end{tabular}
\end{center}
\vspace{2mm}

\section{Relationality in quantum mechanics}

Taking relations as fundamental in quantum mechanics implies a change in the ontology that does not prevent being a realist. The relational interpretation is a realistic one: when a self-adjoint operator, which codes the physical properties of a system, assumes a certain eigenvalue, this corresponds to an element of reality.  Rovelli refers to these elements of reality of relational quantum mechanics as {\em facts} (see Rovelli's article in this volume), or {\em events}.  Like the events in relativity, quantum events are about physical systems in interaction. We may label these systems as {\em observer} or ``observed'', but subjectivity, agents, mind, idealism, phenomenology, etcetera, play no role in this interpretation.  

These facts actually have a clear correspondence with the ontology of classical mechanics. In the  quantum formalism, the {\em observables} correspond by definition to the measurable quantities of classical physics. On the other hand, the relational interpretation is characterized by the fact that these {\em facts} are understood as intrinsically relational: they are real, but their actualization as real is always related to both systems interacting when the value of a measurable quantity is determined.  A fact can be true or actualized with respect to a system (which acts with the abstract role of observer or measuring apparatus) and also not be true with respect to another.

Reality and relationality are therefore inextricably linked.  We attribute existence to a system from its possessing certain properties:  location, speed, energy,  angular momentum,  area, volume...   In quantum mechanics, we realize that it is not it is possible to speak of any of these properties except in a relational manner.  Each property is determined by a relationship between systems. When this relationship does not materialize, the property is not determined. 

In a Galilean system, in order to define the speed of an object we must have a reference system with respect to which the speed is measured; different reference systems associate a different speed to the same object. If no reference system is defined, the object does not have any ``speed'' property. There can not be a notion of ``speed'', for instance, associated to the universe as a whole. 

In relational quantum mechanics this principle is the foundation of the ontology of the theory: the elements of reality, the facts are aspects of a relationship, and take place in interactions. The ontological priority of the interactions invests the whole structure of what we call real%
\footnote{
Notice that in an ontology of relations it is still possible to refer to {\em relata} in a meaningful way. For instance here we have used the notion of systems and we will talk about objects such as particles and fields. All of them, it is argued, have a relational nature.  But, as a structuralist would say, it does not follow from logical principles that they cannot be objects of predication \cite{Saunders:2003}.
}. 

In particular, interactions determine our notion of locality. In contemporary physics we emphasize the fact that interactions are local.  But the notion of interaction is more primitive than that of localization. Nonetheless, as we shall see, it is precisely the locality of the interactions that saves us from some apparent paradoxes of quantum mechanics.

The prototype of these paradoxes is the EPR one \cite{
Smerlak:2006gi,Martin-Dussaud:2018kmh%
}. Two spatially separated systems ({\em observers}) A and B interact with ({\em measure}) two entangled particles, one each.   This determines a fact relative to A and a fact relative to B. A paradox arises only from the assumption that what is an element of reality for A is also an element of reality for B, and vice versa.  A and B may have an element of reality in common only when a local interaction occurs between them.  We cannot consider a fact for A, or a fact for B, as absolute. We can eventually introduce a third observer C, with respect to which there will be some element that regards the comparison of the two, but only provided that this is interacting (hence locally interacting) with both A and B.

\section{The relationality of symmetries}

In modern physics, interactions are largely encoded in symmetries. The symmetries of a system determine the possible interactions such system can have with another system. Symmetries therefore capture the potential of interactions among systems. The apparent arbitrariness that often appear in the definition of the symmetries of a system reflect different possibilities for the system to couple to another system.  For instance, general relativity formulated in a tetrad formalism has an additional local symmetry with respect to Einstein's metric formulation, which captures the possibility of coupling the theory to oriented local objects such as fermions. 

In particular, from this perspective, gauge symmetries do not represent redundant superstructures. They do not just express an indeterminism that needs to be eliminated to get a deterministic theory, or a redundancy in any other sense.  The apparent arbitrariness has its origin in the relationship between gauge and relationality.  A gauge transformation is a mathematical redundancy only when we consider systems in isolation.  The coupling of the system with other systems can well be given by (gauge invariant) interactions that couple {\em gauge degrees of freedom} of one system with {\em gauge degrees of freedom} of the other.  Together, new physical degrees of freedom are born in this coupling.  For instance, the Maxwell potential is redundant in the dynamics of the electromagnetic field alone, but is needed in coupling the field to some charged fields such as an electron. The Maxwell potential is not just a redundant mathematical addition to reality: it is the handle through which the electromagnetic field couples to electrons. 

Notice that what is relevant, what captures the essence of physical reality, is the coupling between systems, not what we identify as the system. 
Two systems coupled to each other cancel their respective gauge redundancies: by coupling a gauge-dependent quantity of one system to a gauge-dependent quantity of the other system, we give rise to a give-independent physical interaction.

This observation leads to  an important distinction regarding observable quantities. We refer to gauge-dependent quantities as {\em partial observables} \cite{Rovelli:2001bz}, and gauge-independent quantities as complete observables in the sense of Dirac. Both kinds of quantities are associated with operators whose eigenvalues corresponds to elements of reality. Both are associated to relative facts. In this sense, partial observables and complete observables have the same ontological status. The difference, on the other hand, is clear cut: partial observables can be measured, but cannot be predicted by dynamic equations alone, while gauge-independent observables can be measured and also predicted \cite{Rovelli:2013bf}. In this sense, as Dirac noted, only the latter lead to a determinism in the theory.  The indeterminacy of the evolution of the former simply reflects the fact that their value depends on the dynamical evolution of another system whose equations of motion are not considered.    For instance, the Einstein equations do not determine uniquely the evolution of the metric tensor because a measurement of this tensor is always relative to a specific (say, material) reference frame, whose equations of motion are in general not included in the Einstein's equations alone.

\section{Relationality in quantum field theory}

A striking example of relationality is provided by the notion of particle in quantum field theory. While some presentations of quantum field theory rely heavily on the notion of particle taken as fundamental \cite{Weinberg}, it is also very well known that the number of particles present in a given quantum field theory state depends on the reference system. On a generic curved spacetime, in particular, there is no unique notion of number of particles.   Physically, different particle detectors count different numbers of particles.  Mathematically, in the absence of global Poincar\'e invariance there is no natural Fock structure in the (nevertheless well defined) Hilbert space of states.  

The existence of particle can be true with respect to one system but not with respect to another.  Different detectors probe different bases in the same Hilbert space.  When a detector measure a certain number of particles, we say that the existence of these particles is an element of reality. But the point above makes clear that this is a relational reality: it is the number of particles {\em with respect to the interaction with that detector}. 

 \vskip1em

\section{Relationality in general relativity
} 

The relational nature of space and time has been longly debated. General relativity, while defining space and time as manifestation of the gravitational field, has a structure that is deeply relational \cite{Vidotto:2013qf}. Dynamical objects are not localized with respect to  a fixed background but only with respect to one another.  Notice how the collection of dynamical objects includes the gravitational field itself.  The very structure of spacetime is built upon contiguity relations, namely the property of spacetime regions being ``next to one another". But in the case of the gravitational field, saying that different regions are contiguous one another through their boundaries means exactly that these regions are interacting.

Alternatively, when we couple general relativity to the matter of a material reference system, the components of the gravitational field with respect to the directions defined by this system are gauge-invariant quantities of the coupled system; but they are gauge-dependent quantities of the gravitational field, measured with respect to a given external frame.  In this case, a prototypical example of a partial observable is time: a quantity that we routinely determine (looking at a clock) but we can not predict from the dynamics of the system. 

\section{Relationality in Quantum Gravity}

The relational aspect of spatio-temporal localization that characterizes general relativity and the relational aspect of quantum mechanics that is emphasized by its relational interpretation combine surprisingly well precisely thanks to the fact that interactions are local. This combination provides a solid conceptual structure for quantum gravity \cite{Vidotto:2013qf}. 

In fact, locality is a main discovery of XX century modern physics: interactions at distance of the Newton's kind don't seem to be part of our world. They are only approximate descriptions of reality.   In the particles' standard model, as well as in general relativity, things can interact only when they ``touch'': all interactions are local.  This means that objects in interactions should be in the same place: interaction require localization and localization requires interaction.  To be in interaction correspond to be adjacent in spacetime and vice versa: the two reduce to one another. 

In other words, the fact that interaction are  local means that they require spacetime contiguity. But the contrary is also true: the only way to ascertain that two objects are contiguous is by means of having them interact. 

Therefore we can identify the {\em Heisenberg cut} that defines the separation with respect to which (relative) facts are realized in quantum theory, with the boundary of spacetime regions that define the (relative) localization in general relativity. 

\begin{table*}[h]
\begin{center}
\begin{tabular}{ccc}
{\bf Quantum relationalism}          & {}\hskip3cm{} &         {\bf Einstein's relationalism} \\
  Systems interact     with other systems               & $\longleftrightarrow$&                Systems are located    wrt other systems\\
            Interaction $\Rightarrow$ Localization     &$\longleftrightarrow$ &   
            Localization $\Rightarrow$  Interaction
\end{tabular}
\end{center}
\label{default}
\end{table*}%

By bringing the two perspectives together, we obtain the boundary formulation of quantum gravity \cite{Oeckl:2003vu,Oeckl:2005bv}: the theory describes processes and their interactions. The manner a process affects another is described by the Hilbert state associated to its boundary. The probabilities of one or another outcome are given by the transition amplitudes associated to the bulk, and obtained from the matrix elements of the projector on the solutions of the Wheeler-DeWitt equation. 

Let us make this more concrete. Consider a process such as the scattering of some particles at CERN. If we want to take into account the gravitational field, we need to include it as part of the system. In doing quantum gravity, the gravitational field (or spacetime) is part of the system.
Distance and time measurements are field measurements like the others in general relativity:  they are part of the boundary data of the problem.

Thinking in terms of functional integrals, we have to sum over all possible histories, but also all possible geometries associated to a given finite spacetime region.

To computate a transition amplitude, we fix the boundary data of the process. In a scattering process, these can be the positions of the particles at initial and final times. These positions are defined by rods and clocks. These measure geometrical informations, and geometrical information is given by the gravitational field. The transition amplitudes depend on the value of all fields on the boundary, including the gravitational fields. They do not depend on further variables such as a time and position. These are coded in the boundary gravitational field, which has the information about how much time have lapsed and the distances between the particles. Geometrical and temporal data are encoded in the boundary state, because these include the state of the gravitational field, which is the state of spacetime. \\[1em]
\begin{center}
\includegraphics[height=50mm]{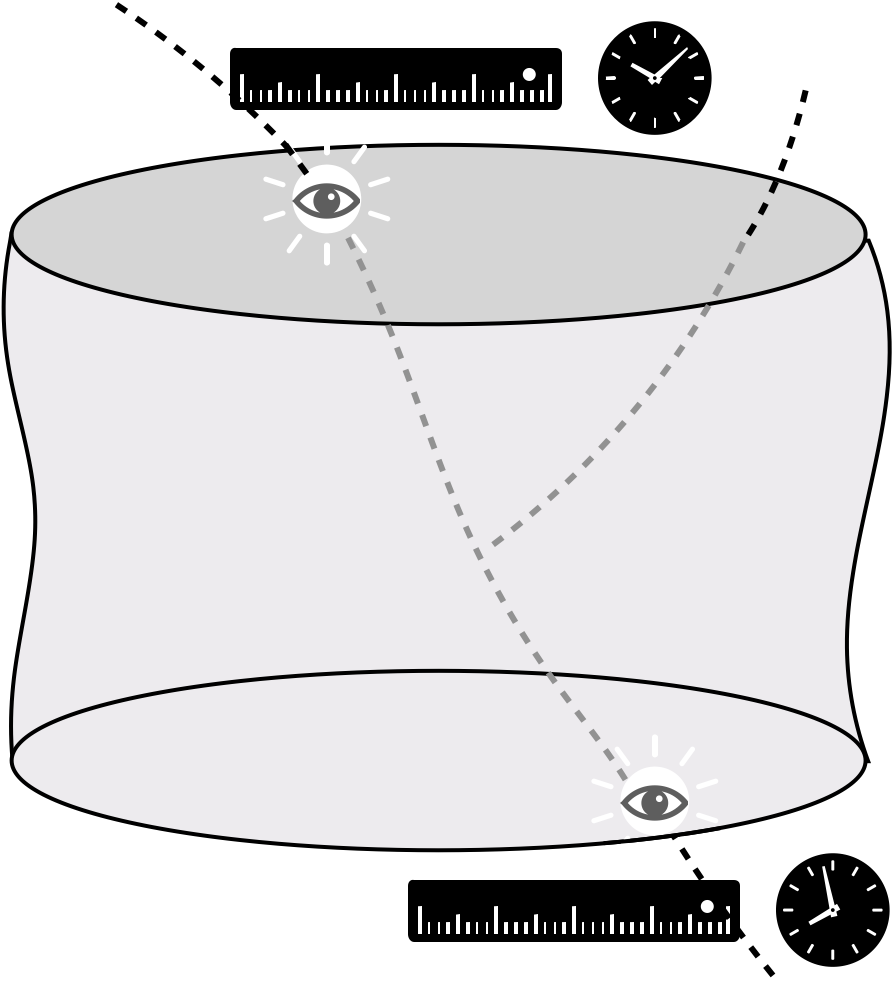} \\
\end{center}

This structural identification is in fact much deeper.  As noticed, the most remarkable aspect of quantum theory is that the boundary between processes can be moved at wish.  Final total amplitudes are not affected by displacing the boundary between ``observed system'' and ``observing system''. The same is true for spacetime: boundaries are arbitrarily drawn in spacetime.  The physical theory is therefore a description of how arbitrary partitions of nature affect one another. Because of locality and because of gravity, these partitions are at the same time subsystems split and partitions of spacetime. A spacetime is a process, a state is what happens at its boundary \cite{Rovelli:2014ssa}. 

This clarifies that in quantum gravity a process is a spacetime region.
Relational quantum mechanics describes systems in interaction. What defines the system and when is it interacting? For spacetime, a process is simply a region of spacetime. Spacetime is a quantum mechanical process once we do quantum gravity. This now helps us to understand how to do quantum gravity.  

Notice that from this perspective quantum gravitational processes are defined locally, without any need to invoke asymptotic regions. Summarizing: 

{\small \begin{center}
{\bf Spacetime Quantum Dynamics}
~~~~~~~~~~~~~~~~~~~~~~~~
\\[1mm]
\begin{tabular}{rcl}
\hline \\
Processes    &~~~~~~~~~~~~~~~~~~ $\longrightarrow$ ~~~~~~~~~~~~~~~~~~ &   Spacetime Regions\\[1em]
States    &$\longrightarrow$ &  Boundaries   (Spacial Regions) \\[1em]
Probability    &$\longrightarrow$&   Transition Amplitudes\\[1em]
Discreteness   &$\longrightarrow$&  Spacetime Quanta \\[1em]
\hline \\ \\
\end{tabular}
\end{center}
\label{default}}


\section{Conclusion: The relational nature of contemporary physics}

The debate on the interpretation of quantum mechanics is far from having reached a consensus. Addressing it is unavoidable in order to answer the question of "what does exist?" as far as current physics tells us. 

But considering this a question related to Quantum Mechanics alone deprives ourselves of some fundamental conceptual inputs, that come from the core of the picture of the world revealed to contemporary physics. 

I have described  the lesson of quantum mechanics from the perspective of relational quantum mechanics. General relativity, quantum field theory and quantum gravity, are compatible and they support such point of view.

Gauge theories and quantum fields theories have a deep relational core: gauge degrees of freedom are handle for interactions to other systems. Even the particles of quantum field theory, that in an ontology of objects we would be tempted to call fundamental objects, are in fact relative, not absolute, entities. 

Locality reveals a deep structural analogy between the relations on which quantum mechanics is based and those on which spacetime is based.  Quantum gravity makes this connection completely explicit. In quantum gravity a process is not in a spacetime region: a process \emph{is} a spacetime region.  Analogously, a state is not somewhere in space: it \emph{is} the description of the way two processes interact, or two spacetime regions pass information to one another. Viceversa, a spacetime region \emph{is} a process: it is like a Feynman sum of everything that can happen between its boundaries. \vspace{2mm}

The resulting relational ontology, compatible with quantum mechanics as well as with the rest of our current physical theories, is a minimalistic one. There is no necessity to attribute an ontological role to states nor some mysterious hidden variables: only facts, or events, are part of the ontology.  It is also a ``lighter'' ontology: facts are \emph{sparse} and \emph{relative}.

This means for instance that particles only exists in interactions, not in between, and exists only with respect to the system they are in relation with, not with respect to the rest of the universe. One may ask: what happens \emph{between} two interactions?
In between, there are other interactions of the field: these interactions are what gives sense to the expression ``in between''. We can distinguish a particle that appears here and then there, being some interaction made by the field: what does define the identity of the particle and its story? Only regularities in the interactions. In fact we may think, if we wish, that there is no particle, only correlated interactions \cite{Hume}. These correlations are such that I measure the field here now and later on there, I obtain correlated values. This is what we mean by saying that there is the same particle. There are just manifestations of a field. A field exists trough its interactions.

This stance weakens \emph{usual} realism, 
but makes it compatible with our current empirical knowledge and spares us pernicious paradoxes.  The relational realism, it should be stressed, is not in any form relativist: going relational does not weaken the reality of the world. If there are only interactions that are intrinsically relational, there is no absolute reality with respect to which the relational events are ``less real''.

Relationalism should not be confused here with a form of subjectivism, which can lead to solipsism. The relations we considered are among any physical systems in interactions, not subjects or agents that require conscious agency. Conscious agents are a peculiar case among the different systems. Systems can acquire and store information about one another: here information should be understood as physical correlations, without a necessary epistemic connotation. 

This leads us to think of relations in a completely physical way, discarding a possible reading of the restriction to the relations as only epistemically motivated (as, for instance, in epistemic structural realism). An interpretation of relations that restricts them to be only epistemic would require the assumption of a hypothetical non-relational underlying substance, not accessible to our knowledge: such a move seems circular and redundant, not adding any clarity to our understanding of the world. In particular, for the sake of philosophy of science, it appears as a useless epicycle.

On the other hand, embracing a relational perspective, we may be able to leave a monolithic reality for a richer kaleidoscopic one. One in which it is required an epistemology where the notion of objectivity is pluralistic and perspectival \cite{Barad:2007,Massimi:2022}.



\end{document}